\documentclass[ArXiv,AXISWP,accept,moreauthors,pdftex,10pt,letterpaper]{Definitions/axis} 

\firstpage{1} 
\pubvolume{xx}
\issuenum{1}
\articlenumber{5}
\pubyear{2023}
\copyrightyear{2023}

\newcommand{\be}{\begin{equation}}
\newcommand{\ee}{\end{equation}}

\newcommand{\msun}{{$M_{\odot}$}}
\newcommand{\mstar}{{$M_{\rm \star}$}}

\newcommand{\ledd}{$L_{\rm Edd}$}

\newcommand{\gtsima}{$\; \buildrel > \over \sim \;$}
\newcommand{\ltsima}{$\; \buildrel < \over \sim \;$}
\newcommand{\prosima}{$\; \buildrel \propto \over \sim \;$}
\newcommand{\gsim}{\lower.5ex\hbox{\gtsima}}
\newcommand{\lsim}{\lower.5ex\hbox{\ltsima}}
\newcommand{\simgt}{\lower.5ex\hbox{\gtsima}}
\newcommand{\simlt}{\lower.5ex\hbox{\ltsima}}
\newcommand{\simpr}{\lower.5ex\hbox{\prosima}}

\newcommand{\lx}{$L_{\rm X}$}

\Title{The black hole occupation fraction of local dwarf galaxies with AXIS}

\Author{Elena Gallo$^{1}$, Edmund Hodges-Kluck $^{2}$, 
Tommaso Treu$^{3}$,
Vivienne Baldassare$^{4}$,
Anil Seth$^{5}$,
Jenny Greene$^{6}$,
Fabio Pacucci$^{7,8}$,
Richard Plotkin$^{9}$,
Amy Reines$^{10}$,
Belinda Wilkes$^{7}$
}

\address{%
$^{1}$ \quad Department of Astronomy, University of Michigan, Ann Arbor\\
$^{2}$ \quad NASA Goddard Space Flight Center\\
$^{3}$ \quad Department of Physics and Astronomy, University of California, Los Angeles\\
$^{4}$ \quad Department of Physics and Astronomy, Washington State University\\
$^{5}$ \quad Department of Physics and Astronomy, University of Utah\\
$^{6}$ \quad Department of Astrophysical Sciences, Princeton\\
$^{7}$ \quad Center for Astrophysics, Smithsonian Astrophysical Observatory\\
$^{8}$ \quad Black Hole Initiative at Harvard University\\
$^{9}$ \quad Department of Physics, University of Nevada, Reno\\
}

\abstract{The fraction of local dwarf galaxies that hosts massive black holes is arguably the cleanest diagnostic of the dominant seed formation mechanism of today's supermassive black holes.  A $\sim$5\% constraint on this quantity can be achieved with AXIS observations of 3,300 galaxies across the mass spectrum through a combination of serendipitous extra-galactic fields plus a dedicated $\sim$1 Msec GO program. \emph{This White Paper is part of a series commissioned for the AXIS Probe Concept Mission; additional AXIS White Papers can be found at the  \href{http://axis.astro.umd.edu/}{AXIS website} with a mission overview \href{https://arxiv.org/abs/2311.00780}{here}}.}

\begin{document}

\tableofcontents
\listoffigures

\section{Massive Black Holes in Dwarf Galaxies: Peeking into the Seeds?}

Massive black holes (BHs) lurking at the center of {nearby} galaxies tend to accrete/shine well below their Eddington limit \cite{Narayan_1994, Narayan_2008, Yuan_Narayan_2014}; as an example, the Galactic Center BH Sgr A* has a bolometric, accretion-powered luminosity of $\sim 10^{-8}$ \ledd \cite{Yuan_2003}. This black hole would be virtually undetectable if it were at the center of the Andromeda galaxy. 
Dynamical methods -- tracing the motion of stars and gas in the vicinity of the BH -- bypass this limitation, and indicate that nearly all Local Volume galaxies as large as the Milky Way, or larger, host a massive BH \cite{McConnell_2012, kormendyho}.  We still do not know if the same is true for ``dwarf" galaxies, with stellar masses below $M_{\star}\simlt 10^9$ and expected BH masses below $M_{\bullet}\simlt 10^5$ \msun \cite{greene20, Pacucci_2021}. Owing to the small sphere of influence, dynamical evidence for a BH can only be acquired in a handful of such galaxies within the Local Volume \cite{McConnell_2012, nguyen19,greene20}. Yet, quantifying the active fraction -- and more importantly, the true BH occupation fraction (hereafter, the BH fraction) -- in dwarf galaxies is crucial to a number of high-stakes astrophysical problems \cite{Bellovary_2011, ricarte18, greene20, Mezcua_2023}. 

\begin{figure*}[t!]
\center{
\includegraphics[width=0.65\textwidth]{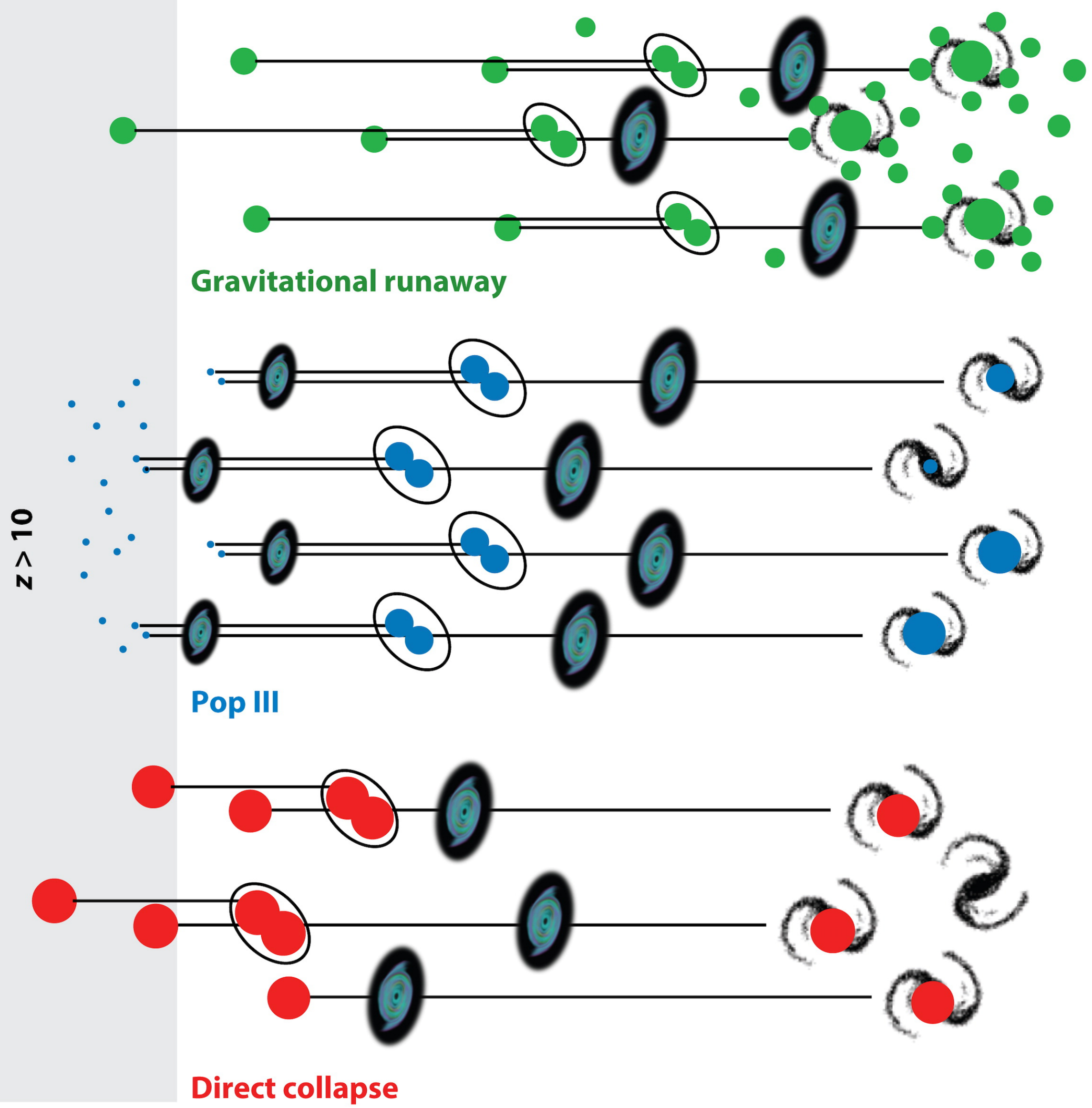}\includegraphics[width=0.285\textwidth]{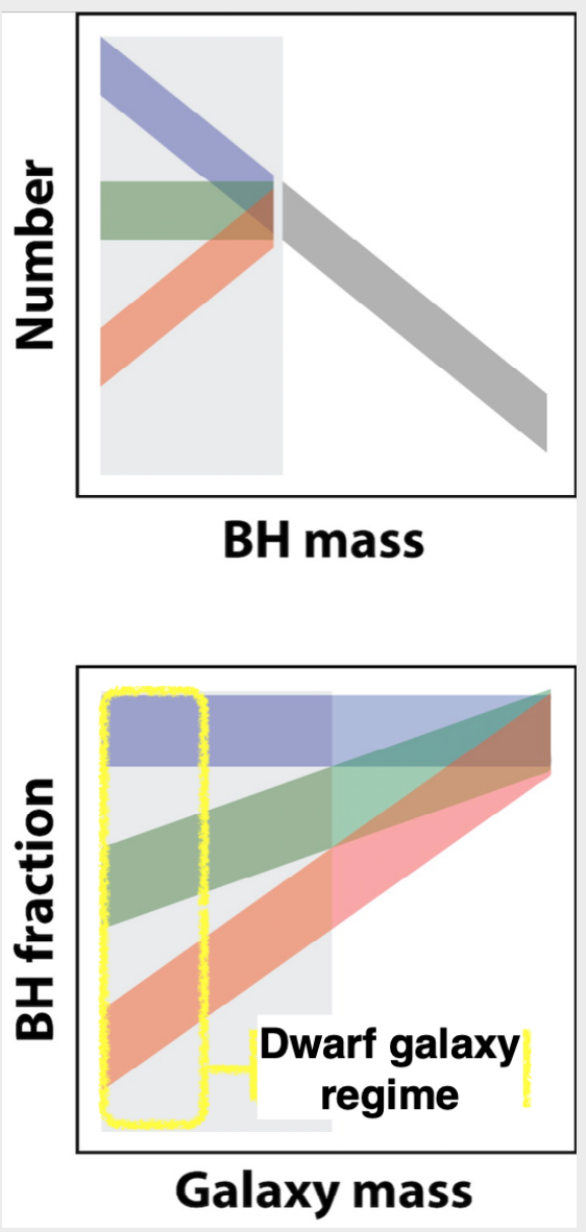}
\caption[Local Black Hole Seeds and Galaxy Mass]{
The local BH fraction in dwarf galaxies is thought to be sensitive to the dominant BH seeding mechanism at $z \simgt 10$. The predictions vary significantly for lower mass galaxies at $z =0$, particularly in the dwarf regime below $\simlt 10^9$ \msun, highlighted in yellow in the bottom-right panel. If the seeds were of the light variety, i.e., POP III star remnants (blue symbols/regions) then virtually \textit{all} local dwarf galaxies will have a massive BH by $z=$0. Conversely, if the primary seeding mechanism was direct collapse of heavy (red), dwarf galaxies at $z=0$ will be devoid of a massive BH. If BHs were seeded via gravitational runaway in dense stellar clusters (green), this would result in intermediate $\sim$50\% BH fraction for $z=0$ dwarfs. This local approach provides an independent constraint to searches for high-$z$ quasars. Adapted from \cite{greene20}. \label{fig:jenny}}}
\rule{\textwidth}{0.4pt}
\end{figure*}

For one, the BH fraction in dwarf galaxies is thought to depend sensitively on the black hole ``seeding" mechanism at high-$z$~(see Fig.~\ref{fig:jenny}, and, e.g., \cite{greene20, Haidar_2022}). At high galaxy masses (Milky Way and higher), the BH fraction is expected to be close to 100\% regardless of the seeding mechanism. Models suggest that by $z$ = 0, virtually all dwarf galaxies will contain a massive BH if stellar-sized Pop III remnants ($M_{\bullet}\simeq 10^2$ \msun) provide the {dominant} seeding mode, as opposed to global gas collapse on galaxy-size scales ($M_{\bullet}\simeq 10^{5-6}$ \msun), which leads to virtually zero BH fractions in dwarfs. Alternatively, if seeding occurs preferentially via gravitational runaway ($M_{\bullet}\simeq 10^{3-4}$ \msun, see, e.g., \cite{PZ_2002, Boekholt_2018, Das_2021}), then about 50\% of local dwarfs can be expected to host a massive BH \cite{volonteri12,natarajan14,agarwal16,valiante16,ricarte18,woods19,ina21}. Additionally, {\it if} the BH fraction of dwarf galaxies were close to 100\%, then their black holes (rather than supernovae) could be entirely responsible for quenching star formation in these systems \cite{silk18,dickey19}. Finally, the expected rates of extreme mass ratio inspirals \cite{babak} and tidal disruption events \cite{stonemetzger} that will be detectable by LISA and Rubin, respectively, are extremely sensitive to the BH fraction in dwarfs. \smallskip\\

Ground-based methods are unable to measure the BH fraction in the dwarf galaxy regime with current 10-m class facilities, but high-resolution X-ray imaging can do it \cite{hk20}. This is because (i) X-rays are a very clean probe of accretion-powered emission, and even formally `inactive' BHs glow due to weak accretion, and (ii) a statistical method exists to reliably convert the \textit{measured} active fraction into BH fraction \cite{miller15,gallosesana}.\smallskip\\

Pursuing a ``local" approach to the BH seeding conundrum is especially compelling, since it is orthogonal and complementary to ongoing and future surveys at high-$z$.
In the high-$z$ Universe, the detection of a massive quasar ($M_{\bullet} > 10^9$ \msun) at a $z\simeq 10$ would rule out a wide parameter space of seed masses for a {\em specific} system \cite{pacucci22,natarajan23}. However, progress towards establishing the {\em dominant} BH seeding mechanism requires next-generation X-ray (and IR) facilities, as only the wide field of view. and $\sim$arcsec resolution afforded by a Chandra successor mission, such as AXIS, will be able to detect smaller/less luminous BHs beyond $z\simgt 7$ \cite{hk20,civano,haiman}.  

\section{How AXIS can Measure the BH Occupation Fraction}

{ Although Chandra has shown conclusively that the occupation fraction is $>$50\% down to galaxy stellar masses \mstar$\simeq 10^{9}$ \msun\ \cite{greene20}, higher precision is needed to constrain seeding models and other processes that impact occupation. Here, we articulate a strategy to deliver a 5\% measurement of the local BH occupation fraction, including extending measurements to \mstar $< 10^9$ \msun\ that are inaccessible with Chandra. }

\subsection{Counting ``Missing'' Black Holes with AXIS}

Within the Local Volume, hard X-rays are the best signpost of massive BHs because the sky is dark, and even `quiescent' BHs emit X-rays due to persistent, weak accretion from the interstellar medium or stellar mass loss \cite{pellegrini10,miller12,she17,bi20}. Moreover, we know from Chandra that these BHs have predictable luminosities that depend on the stellar mass: there is a positive, log-linear relationship between the X-ray luminosity (\lx) and \mstar\ measured by \cite{miller12,miller15}. This relation is unsurprising because larger BHs inhabit larger galaxies and are more luminous at a given Eddington fraction. There is also a significant intrinsic scatter ($\simeq$0.9 dex) that is also expected and explained by differences in accretion rates, local absorption effects, and scatter in BH mass at a fixed \mstar. Importantly, at least down to $\log$~\mstar $=8$, \lx\ is never {\it zero}.

This means that a perfect X-ray telescope would always detect a massive BH in each galaxy at full occupation. In other words, the detected fraction {\it is} the occupation fraction. But even an imperfect telescope can constrain the occupation fraction by leveraging the \lx:\mstar\ relation. At a given \mstar, the \lx:\mstar\ relation and its scatter predict the detection fraction at a given sensitivity. By specifying a precision on the measurement of this fraction, we obtain the number of galaxies that must be observed to rule out full occupation (or any other input occupation fraction exceeding zero). Conversely, we constrain occupation from the low end by detecting more BHs than expected from a too low occupation fraction. 

The sensitivity is clearly the key to making a precise measurement feasible, since a telescope achieving a sensitivity 1$\sigma$ below the mean will need far fewer galaxies to establish the occupation fraction than one whose sensitivity is 1$\sigma$ above the mean. AXIS has the sensitivity to detect BHs that emit at or below the mean \lx:\mstar\ down to $\log$~\mstar\ $>8$ in the Local Volume and in a reasonable time.

\subsection{X-ray Binary Contamination}

However, this otherwise clean measurement is complicated by the presence of X-ray binaries (XRBs), since XRBs emitting near the Eddington limit ($\simeq 10^{38}$~erg~s$^{-1}$ for one solar mass) can have a similar luminosity as a very weakly accreting, but more massive BH. At very low luminosities, even a collection of sub-Eddington XRBs at or near the galactic nucleus can be mistaken for a massive BH. 

Without a perfect telescope, one cannot determine from X-rays alone whether a given nuclear X-ray source is an XRB or massive BH. However, the number of XRBs scales with \mstar\ (low-mass XRBs) and star-formation rate (high-mass XRBs) and the luminosity functions of these XRBs are well constrained by existing data. Hence, one can predict the odds of detecting nuclear X-rays from XRBs at a given telescope sensitivity, angular resolution, galaxy mass, distance, etc. This involves calculating the mean luminosity expected from stellar mass or star formation enclosed in a nuclear aperture and sampling from the luminosity function, as described in \cite{hk20}. 

The outcome is the probability of detecting XRB emission at or above a given sensitivity for a given galaxy ($P_{XRB}$). $P_{XRB}$ can be combined with the expected detection fraction for massive BHs at a fixed sensitivity to produce the number of galaxies needed to measure the occupation fraction. Higher sensitivity reduces the number of required galaxies, but this must be balanced against $P_{XRB}$. $P_{XRB}$ depends linearly on angular resolution \cite{hk20}, so X-ray telescopes with angular resolution significantly worse than 1~arcsec require prohibitively large sample sizes. 

\subsection{AXIS Performance and Occupation Fraction}

Using the current best estimate of AXIS angular resolution (1.25$^{\prime\prime}$ on-axis and 1.5$^{\prime\prime}$ averaged over a 24-arcmin diameter circular field of view) and effective area (3600 cm$^2$ at 1~keV, averaged over the field of view), we can estimate the number of galaxies needed to measure the local occupation fraction to a precision of 5\%. Following \cite{hk20}, we calculate $P_{XRB}$ as a function of galaxy mass, size, and distance for a catalog of SDSS galaxies within 100~Mpc. $P_{XRB}$ is less than 0.1 for more than 90\% of galaxies with $\log$ \mstar\ $<9.0$, with the exceptions having unusually high star formation or high concentration. For $P_{XRB} < 0.1$, about 1100 galaxies in each 0.5~dex bin from $8.0 < $\mstar $< 9.5$ is sufficient to measure the occupation fraction to about 5\% precision. This optimal number is achieved by adopting sensitivity limits in each mass bin that are a fixed distance away from the mean \lx, based on the scatter from \cite{miller15}. 

This number is quite conservative when considering only $P_{XRB}$, as in the lowest mass bin only $\simeq$500 galaxies would be required based on the estimated $P_{XRB}$ within 100 Mpc. However, our calculation assumes a smooth shape (exponential disk) and no clumpy star formation across the mass scale. Assuming that a maximal $P_{XRB}$ covers over these issues and does not (see below) prevent AXIS from making this measurement.

With its excellent resolution and low instrumental background, AXIS is essentially photon-limited for exposures shorter than several hundred ks. We can then calculate distance horizons for sources of various luminosities from the average effective area across the field of view (3600~cm$^2$ at 1 keV), assuming 5$\sigma$ detection significance in the 0.5-2~keV band. For example, a source with $L_{X} = 10^{38}$~erg~s$^{-1}$ is detectable to 13~Mpc in 10~ks, 29~Mpc in 50~ks, and 53~Mpc in 100~ks, with horizons scaling from these values as the square root of the luminosity.

\subsection{A Commensal AXIS Survey and Modest GO Program}

If we need 3300 galaxies to measure the occupation fraction to 5\% precision, observing them one by one is prohibitively expensive. AXIS will have at least 105~Ms of total science exposure over five years, of which 70\% is available for general observing. Half of this would be required for 10~ks snapshots of 3600 galaxies. However, because most AXIS fields (especially those at high Galactic latitude) will contain a few dwarf galaxies within 100~Mpc by chance, it is possible to exploit the AXIS archive to achieve almost all of this program.

Following \cite{hk20}, who laid out a speculative observing strategy for a flagship-class X-ray mission with subarcsecond resolution, we examine the types of observations that AXIS is likely to make based on the Chandra archive. In addition to the SDSS catalog they used, we also adopt the conclusions of \cite{hk20} that low surface brightness galaxies are legitimate targets and that the SDSS luminosity function implies that there are more than 100,000 galaxies per mass bin within 100~Mpc, across the whole sky. 

Although this is many galaxies, if they are randomly distributed across the sky (0.3 galaxies per mass bin, per AXIS field), this is insufficient to complete the sample. If we assume that the AXIS GO time (73 Ms) is divided into 10, 50, and 100~ks exposures of unique fields and that 25\% of those fields are crowded, in the Galactic plane, or otherwise unusable for this work (based on the Chandra observing log), we do not serendipitously find enough galaxies even in the high mass bin ($9.0 < \log$\mstar$ < 9.5$), where we have the largest distance horizon. We expect about 650 and need 1100. 

However, galaxies are not randomly distributed. Many AXIS observations will target galaxies within 100~Mpc, and galaxies tend to group. Based on a catalog of bright central galaxies \cite{blanton05}, we expect 1-3 galaxies in the required mass range within the AXIS field of view when AXIS targets a central, $L*$ galaxy. Further, based on the proportion of Chandra observations targeting such galaxies, we expect at least 900 serendipitous dwarf galaxies in addition to those caught randomly (i.e., without targeting their central galaxy). Cluster outskirts, a key AXIS topic, will contain a large number of gas-poor, dwarf, low-surface-brightness galaxies where the intracluster medium is dim enough to detect faint nuclear emission that could easily be sustained by stellar mass loss or accreting from the intracluster medium. The so-called ``ultra-diffuse'' galaxies in clusters like Coma \cite{koda15} imply several hundred per nearby cluster in the required mass regime and outside of the bright cluster core. If AXIS surveys Coma, Perseus, and Virgo, this would amount to about 1000 galaxies per mass bin. We then assumed uniformly distributed 0.3 galaxies per bin, per field, within 100 Mpc for every other field. Finally, we assume that 50\% of the galaxies are not suitable because a nucleus cannot be clearly identified. With these assumptions, we expect to exceed 1100 galaxies in the upper two mass bins ($8.5 < \log$\mstar$ < 9.5$) but only obtain 500 or less in the lowest mass bin ($8.0 < \log$\mstar$ < 8.5$). 

This is an optimistic but realistic observing plan (i.e., based on past performance). However, even if this is realized, we will not sample enough low-mass galaxies. A targeted GO program is thus required, but it can be optimized so that we do not need to observe 600 galaxies. Galaxies can be selected on the basis of proximity, lack of nuclear star formation, and other criteria. Moreover, as described above, in the low-mass bin $P_{XRB} \ll 0.1$ for the sensitivity achieved by AXIS in $<$10~ks snapshots, regardless of distance. It is realistic to expect that the total number of additional galaxies will be closer to 200 than 600, and that the total time required will be less than 1.5~Ms. 

Finally, this exercise is based on 5 years of AXIS operations but the observing plan for over 20 years of Chandra operations. If AXIS operates for 15 years at baseline performance, there is almost no doubt that it can serendipitously achieve this measurement.

\vspace{6pt} 




\appendixtitles{no} 
\appendixsections{multiple} 



\externalbibliography{yes}
\bibliography{references}

\end{document}